\documentclass[twocolumn,english,prd,nolongbibliography,superscriptaddress,aps,showpacs,showkeys,amsmath,amssymb,floatfix,nofootinbib]{revtex4}
\usepackage{lmodern}

\usepackage[T1]{fontenc}
\usepackage[utf8]{inputenc}
\usepackage{amsmath}
\usepackage{graphicx}
\usepackage{amssymb}
\usepackage{esint}

\makeatletter

\providecommand{\tabularnewline}{\\}

\@ifundefined{textcolor}{}
{%
 \definecolor{BLACK}{gray}{0}
 \definecolor{WHITE}{gray}{1}
 \definecolor{RED}{rgb}{1,0,0}
 \definecolor{GREEN}{rgb}{0,1,0}
 \definecolor{BLUE}{rgb}{0,0,1}
 \definecolor{CYAN}{cmyk}{1,0,0,0}
 \definecolor{MAGENTA}{cmyk}{0,1,0,0}
 \definecolor{YELLOW}{cmyk}{0,0,1,0}
 }

\@ifundefined{definecolor}
 {\@ifundefined{definecolor}
 {\usepackage{color}}{}
}{}

\pdfoutput=1
\@ifundefined{definecolor}
 {\@ifundefined{definecolor}
 {\@ifundefined{definecolor}
 {\@ifundefined{definecolor}
 {\@ifundefined{definecolor}
 {\usepackage{color}}{}
}{}
}{}
}{}
}{}
\usepackage[colorlinks,bookmarks]{hyperref}\definecolor{linkblue}{rgb}{0,0,0.8}
\definecolor{linkgreen}{rgb}{0,0.5,0}
\hypersetup{linkcolor=linkblue, citecolor=linkgreen, urlcolor=linkblue}

\usepackage{lmodern}\usepackage{babel}%\setcounter{secnumdepth}{3}
\usepackage{bm}\usepackage[us]{datetime}

\providecommand{\tabularnewline}{\\}

\def\rd{{\rm d}}

\makeatother

\usepackage{babel}

\makeatother

\usepackage{babel}

\makeatother

\usepackage{babel}

\makeatother

\usepackage{babel}

\makeatother

\usepackage{babel}

\begin{document}
%%%%%%%%%%%%%%%%%%%%%%%%%%%%%%%%%%%%
%%%%%%%%%%%%%%%%%%%%%%%%%%%%%%%%%%%%

\title{Supernova constraints on Multi-coupled Dark Energy}

\author{Arpine Piloyan}

\affiliation{Yerevan State University, Alex Manoogian 1, Yerevan 0025, Armenia}

\author{Valerio Marra}

\affiliation{Institut für Theoretische Physik, Universität Heidelberg, Philosophenweg
16, 69120 Heidelberg, Germany}

\author{Marco Baldi}

\affiliation{Dipartimento di Fisica e Astronomia, Università di Bologna, Viale
C. Berti-Pichat 6/2, I-40127, Bologna}

\affiliation{ INAF, Osservatorio Astronomico di Bologna, Viale C. Berti-Pichat
6/2, I40127, Bologna}

\author{Luca Amendola}

\affiliation{Institut für Theoretische Physik, Universität Heidelberg, Philosophenweg
16, 69120 Heidelberg, Germany}
\begin{abstract}
The persisting consistency of ever more accurate observational data
with the predictions of the standard $\Lambda$CDM cosmological model
puts severe constraints on possible alternative scenarios, but still
does not shed any light on the fundamental nature of the cosmic dark
sector. As large deviations from a $\Lambda$CDM cosmology are ruled
out by data, the path to detect possible features of alternative models
goes necessarily through the definition of cosmological scenarios
that leave almost unaffected the background and -- to a lesser extent
-- the linear perturbations evolution of the universe. In this context,
the Multi-coupled DE (McDE) model was proposed by Baldi~\cite{Baldi_2012a}
as a particular realization of an interacting Dark Energy field characterized
by an effective screening mechanism capable of suppressing the effects
of the coupling at the background and linear perturbation level. In
the present paper, for the first time, we challenge the McDE scenario
through a direct comparison with real data, in particular with the
luminosity distance of Type Ia supernovae. By studying the existence
and stability conditions of the critical points of the associated
background dynamical system, we select only the cosmologically consistent
solutions, and confront their background expansion history with data.
Confirming previous qualitative results, the McDE scenario appears
to be fully consistent with the adopted sample of Type Ia supernovae,
even for coupling values corresponding to an associated scalar fifth-force
about four orders of magnitude stronger than standard gravity. Our
analysis demonstrates the effectiveness of the McDE background screening,
and shows some new non-trivial asymptotic solutions for the future
evolution of the universe. Clearly, linear perturbation data and,
even more, nonlinear structure formation properties are expected to
put much tighter constraints on the allowed coupling range. Nonetheless,
our results show how the background expansion history might be highly
insensitive to the fundamental nature and to the internal complexity
of the dark sector.
\end{abstract}



\maketitle
%\textbf{\color{red} \today; \currenttime}

\section{Introduction}

The standard model of cosmology -- characterized by the existence
of two distinct forms of gravitating energy that do not interact with
the electromagnetic field, thereby evading direct observations --
has been tremendously challenged over the past decade by the impressive
improvements in the accuracy of observational tests. Despite the wide
variety of complementary probes that have been progressively developed
in order to test the consistency of the model, covering a huge range
of scales and redshifts, no significant deviation from the predictions
of standard cosmology has been detected so far, even with the exquisite
accuracy recently reached by the Planck satellite \citep{Planck_016}.
Nonetheless, such ever-increasing accuracy in the consistency checks
on the model and in the precise determination of its few basic parameters
has shed no light on the fundamental nature of the two dominating
constituents of the Universe, Dark Energy (DE) and Cold Dark Matter
(CDM).

While the former appears to be fully consistent with the predictions
of a cosmological constant $\Lambda$, more sophisticated possibilities
like dynamical or interacting DE, or large-scale modifications of
gravity (see e.g. \cite{2010deto.book.....A}), have not yet been excluded, and retain most of their original
appeal as possibly alleviating the fundamental fine-tuning problems
of $\Lambda$. In particular, several alternative DE scenarios can
mimic the evolution of a cosmological constant closely enough to evade
present observational bounds. Similarly, while the evidence in favor
of a CDM component from astrophysical and cosmological observations
is now hardly controvertible, none of the experimental efforts put
in place so far have been able to provide a clear and statistically
significant detection of any of the plausible CDM particle candidates,
arising in different extensions of the standard model of particle
physics.

In such context, it is interesting to explore speculative models with
a higher level of internal complexity of either of these two mysterious
dark fields, or possibly of both, as long as present observational
bounds are matched. In particular, for what concerns the DE sector,
dynamical DE models such as {\em Quintessence} \citep{Wetterich_1988,Ratra_Peebles_1988}
or {\em k-essence} \citep{kessence} have been proposed, possibly
also featuring non-negligible perturbations at sub-horizon scales
\citep{Creminelli_etal_2009,Creminelli_etal_2010,Sefusatti_Vernizzi_2011,Batista_Pace_2013}
or direct interactions with CDM \citep{Damour_Gibbons_Gundlach_1990,Wetterich_1995,Amendola_2000,Baldi_2011a}
or massive neutrinos \citep{Amendola_Baldi_Wetterich_2008}, as well
as modified gravity models \citep{Hu_Sawicki_2007}. While a large
portion of the parameter space of such models has been progressively
ruled out by increasingly accurate observational constraints, the
range of parameters that remains viable still offers the chance of
some relevant non-standard phenomenology \citep[see e.g.][]{Pettorino_etal_2013}.
Similarly, for the CDM sector, several possible extensions of the
standard model have been proposed, ranging from Warm Dark Matter models
\citep[WDM, see e.g.][]{Bode_Ostriker_Turok_2001}
to Self-Interacting Dark Matter \citep[SIDM, see e.g.][]{Loeb_Weiner_2011,Aarssen_Bringmann_Pfrommer_2012,JonathanL.Feng2010, Vogelsberger_Zavala_2013}
or mixed Cold+Warm Dark Matter scenarios \citep{Khlopov_1995,Maccio_etal_2012},
characterized by the existence of more than one Dark Matter particle
species. More recently, models with almost degenerate multiple CDM species have been also considered \citep[see][]{Medvedev_2013}.

While the most basic and widely investigated models of interactions
between the DE field and CDM particles are now robustly constrained
through their predicted impact on the background and linear perturbation
evolution \citep[see e.g.][]{Bean_etal_2008,LaVacca_etal_2009,Xia_2009,Pettorino_etal_2012, 2013arXiv1304.7119} as
well as through their effects on nonlinear structure formation \citep{Maccio_etal_2004,Baldi_etal_2010,Baldi_2011a,Li_Barrow_2011},
\citet{Brookfield_vandeBruck_Hall_2008} have recently shown that
if DE interactions are associated with a higher level of complexity
of the CDM sector, such constraints might significantly relax, leaving
room for a new class of non-standard cosmological scenarios. More
specifically, if DE interacts with different couplings to different
species of CDM particles, the impact of the interaction on background
and linear perturbations can be relatively mild even for individual
coupling values largely exceeding the present observational bounds
for the case of an interaction with a single CDM species.

While the drawback of such class of models might reside in the need
to introduce a large number of new free parameters associated with
each individual interaction channel, we will focus in the present
paper on a particular realization of this scenario recently proposed
by \citet{Baldi_2012a}, which involves only two different species
of CDM particles characterized by the same absolute value of the coupling
to the DE field but with opposite signs, thereby requiring no more
parameters than a standard interacting DE model.

In the present paper, we will provide the first direct comparison
of such a ``Multi-coupled DE'' (McDE) scenario with the supernova
luminosities of the publicly available Union2.1 sample \citep{Suzuki_etal_2012}. Confirming previous qualitative results, our analysis will show how
present observational data on the background expansion are fully consistent
with McDE scenarios even for very large values of the DE coupling,
up to  three orders of magnitude larger than present bounds on the coupling for standard cDE models.

The present work is organized as follows. In section \ref{background}
we will review the main background equations characterizing the McDE
scenario; in section \ref{critical_points} we will perform a phase-space
analysis of the critical points of the system, discussing their existence
and stability conditions; in section \ref{supernovae} we will introduce
the data sample adopted in our analysis and the methods employed to
compare data with the McDE expectations, and in section \ref{results}
we will discuss the outcomes of such comparison. Finally, in section
\ref{conclusions} we will draw our conclusions.

\section{Background equations}

\label{background}

As mentioned above, we will study the model proposed in the Ref.~\cite{Baldi_2012a},
defined by the action integral \begin{align}
S = & \int d^{4}x\sqrt{-g}\biggl[\frac{M_{Pl}^{2}}{2}R+\frac{1}{2}\dot{\phi}^{2}-\underset{\pm}{\sum}m_{\pm}e^{\pm C\phi}\bar{\psi}_{\pm}\psi_{\pm}\nonumber \\
 & -V(\phi)+\mathcal{L}_{r}\biggl]\,,\label{eq:Action-1}\end{align}
 where $g$ is the determinant of the metric tensor, $M_{Pl}=1/\sqrt{8\pi G}$
is the reduced Planck mass and $\mathcal{L}_{r}$ is the Lagrangian
of the relativistic components. The new idea of the model is the introduction
of two different CDM species $\psi_{\pm}$ interacting with the scalar
field $\phi$ through the dimensional couplings $\pm C$. For simplicity,
we will not consider the baryonic component of the Universe in this
work. In a homogeneous and isotropic Friedmann-Lemaître-Robertson-Walker
(FLRW) model with scale factor $a$ and no spatial curvature, the
field equations of McDE cosmologies are as follows:
\begin{align}
\ddot{\phi}+3H\dot{\phi}+\frac{dV}{d\phi} & =+C\rho_{-}-C\rho_{+}\,,\label{eq:F_EQ_phi}\\
\dot{\rho}_{-}+3H\rho_{-} & =-C\,\dot{\phi}\,\rho_{-}\,,\label{eq:F_EQ_1}\\
\dot{\rho}_{+}+3H\rho_{+} & =C\,\dot{\phi}\,\rho_{+}\,,\label{eq:F_EQ_2}\\
\dot{\rho}_{r}+4H\rho_{r} & =0\,,\label{eq:F_EQ_3}\\
3M_{Pl}^{2}\, H^{2} & =\rho_{r}+\rho_{+}+\rho_{-}+\rho_{\phi}\,,\label{eq:F_EQ_4}\end{align}
 where an overdot represents a derivative with respect to the cosmic
time $t$, $H\equiv\dot{a}/a$ is the Hubble function, and the total
CDM density is related to the two dark matter densities $\rho_{\pm}$
by $\rho_{{\rm CDM}}=\rho_{+}+\rho_{-}$. As in Ref.~\cite{Baldi_2012a}
we introduce the asymmetry parameter $\mu$ as follows:
\begin{equation}
\mu=\frac{\Omega_{+}-\Omega_{-}}{\Omega_{+}+\Omega_{-}}\,,\label{eq:Myu}\end{equation}
 where the fractional density parameters $\Omega_{i}$ are given by
\begin{equation}
\Omega_{i}=\frac{\rho_{i}}{3H^{2}M_{Pl}^{2}}\,.\label{eq:Omega}\end{equation}
 We will denote with a subscript {}``0'' and {}``in'', respectively,
present-day and initial values of time-dependent parameters such as
$\mu$. %\begin{equation}
%\mu_{0}=\frac{\Omega_{0+}-\Omega_{0-}}{\Omega_{0+}+\Omega_{0-}}\,,\label{eq:Myu_0}
%\end{equation}
% and
%\begin{equation}
%\mu_{{\rm in}}=\frac{\Omega_{{\rm in+}}-\Omega_{{\rm in-}}}{\Omega_{{\rm in+}}+\Omega_{{\rm in-}}}\,.\label{eq:Myu_in}
%\end{equation}
We also introduce a dimensionless coupling constant $\beta$ defined
as: \begin{equation}
\beta\equiv\sqrt{\frac{3}{2}}M_{Pl}\, C\,.\label{eq:Cp_cst}\end{equation}
 Following Ref.~\cite{Baldi_2012a}, we restrict our analysis to
the case of an exponential self-interaction potential of the form:
\begin{equation}
V(\phi)=V_{0}\, e^{-\alpha_{1}\phi}\,\label{eq:POtential}\end{equation}
 This particular choice simplifies the equations in analogy with the
single-field coupled dark energy model of Refs~\cite{Wetterich_1995,Amendola_2000}.

\begin{figure}
\includegraphics[width=0.95\columnwidth]{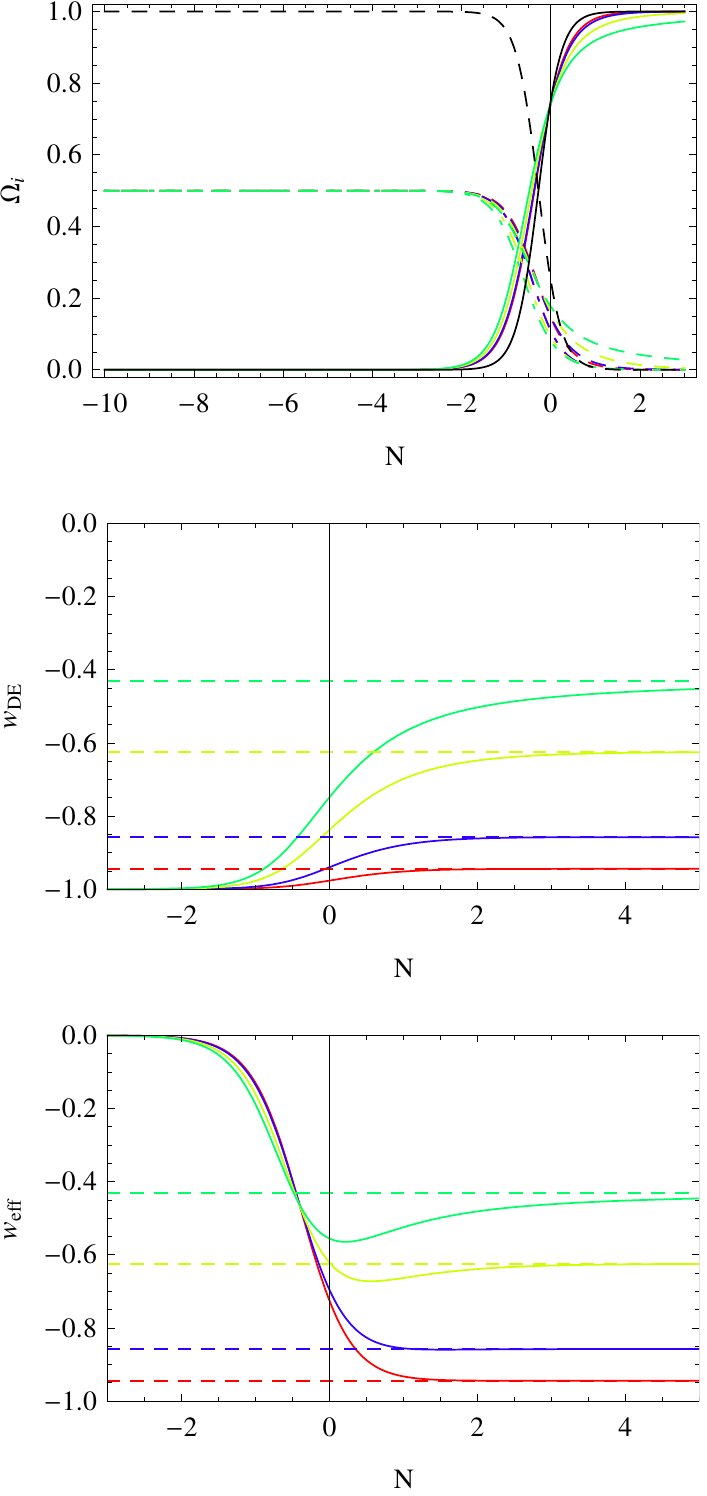}
\caption{In the top panel we show plots of $\Omega_{DE}$, $\Omega_{-}$ (dot-dashed
curve) and $\Omega_{+}$ (dashed curve) for the case of coupling
$\beta=1$ and $\mu_{{\rm in}}=0$, when (from bottom up) $\alpha=0.5$
(red curve), $\alpha=0.8$ (blue curve), $\alpha=1.3$ (yellow curve)
and $\alpha=1.6$ (green curve). These values of $\alpha$ belong
to the $1\sigma$, $2\sigma$, $3\sigma$ and $>3\sigma$ regions
of the last panel of Fig.~\ref{2dpostP2}, respectively. The black curves correspond to the $\Lambda CDM$ model, where $\Omega_{\Lambda}$ is the solid black curve and  $\Omega_{m}$ is the dashed black curve. The second and third panels
show plots of $w_{DE}$ and $w_{eff}$ for the same parameters with
the same color coding. The horizontal dashed curves in the latter
two panels are the corresponding theoretical predictions for the asymptotic
behavior, again with the same color coding. All plots are with respect
to the e-folding time variable $N\equiv\ln a$, where $a$ is the
scale factor. See Section \ref{background} for more details.
Note that vertical lines represent the present time, which corresponds to $N=0$.}
\label{p_1mu_0} %
\end{figure}

\begin{figure}
\includegraphics[width=0.95\columnwidth]{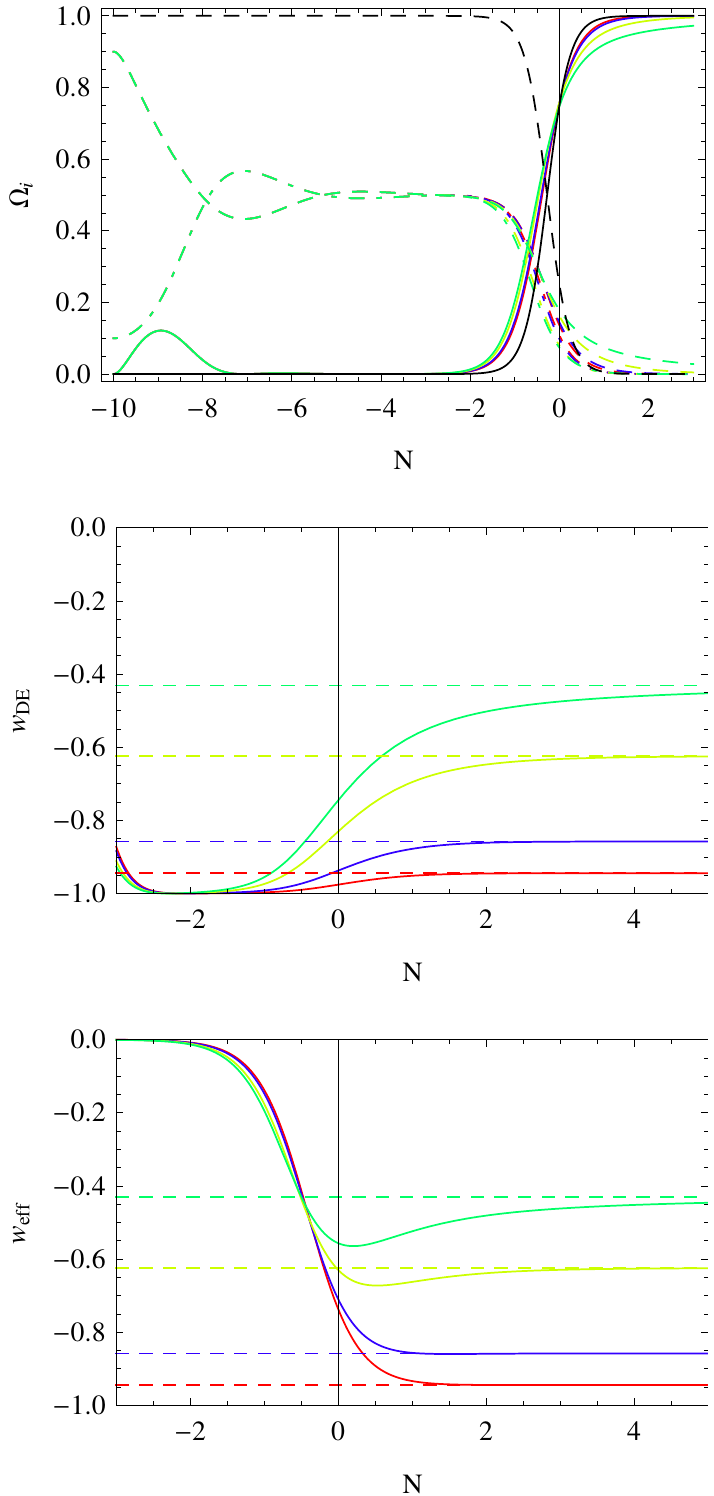}
\caption{As in Fig.~\ref{p_1mu_0} but for $\mu_{{\rm in}}=0.8$.}
\label{p_1mu_08} %
\end{figure}

Let us now introduce the following notation: \begin{align}
\frac{\dot{\phi}^{2}}{6M_{{\rm Pl}}^{2}H^{2}} & =x^{2}\,,\label{eq:Cont_Sc}\\
\frac{V}{3M_{{\rm Pl}}^{2}H^{2}} & =y^{2}\,,\\
\frac{\rho_{\pm}}{3M_{{\rm Pl}}^{2}H^{2}} & =z_{1,2}^{2}\,,\\
\frac{\rho_{r}}{3M_{{\rm Pl}}^{2}H^{2}} & =r^{2}\,,\label{eq:Cont_Sc_last}\end{align}
 After replacing the new variables (\ref{eq:Cont_Sc}-\ref{eq:Cont_Sc_last})
and introducing the new e-folding time variable $N\equiv\ln a$ (we
will denote the derivative with respect to $N$ with a prime) and
after defining
\begin{equation}
\alpha=\sqrt{\frac{3}{2}}M_{Pl}\,\alpha_{1}\,,\label{eq:Cp_cst-1}
\end{equation}
we obtain the full set of phase-space equations of McDE cosmologies:
\begin{align}
2\frac{H'}{H} & =-(3-3y^{2}+3x^{2}+r^{2})\,,\label{eq:Frid_EQ-H-prime}\\
x' & =-x\frac{H'}{H}-3x+\alpha y^{2}+\beta(z_{2}^{2}-z_{1}^{2})\,,\label{eq:Frid_EQ-1}\\
y' & =-y\frac{H'}{H}-\alpha xy\,,\label{eq:Frid_EQ-2}\\
z_{1}' & =-z_{1}\frac{H'}{H}-\frac{3}{2}z_{1}+\beta xz_{1}\,,\label{eq:Frid_EQ-3}\\
z_{2}' & =-z_{2}\frac{H'}{H}-\frac{3}{2}z_{2}-\beta xz_{2}\,,\label{eq:Frid_EQ-4}\\
r' & =-r\frac{H'}{H}-2r\,.\label{eq:Frid_EQ-5}\end{align}
The former system of field equations (\ref{eq:F_EQ_phi})-(\ref{eq:F_EQ_4})
are symmetric for opposite signs of $y$, $z_{1}$, $z_{2}$ and $r$.
However, only the square of these quantities has physical meaning
and there is no need to consider solutions of both signs. Notice that
the dark energy density is
\begin{equation}
\Omega_{{\rm DE}}=x^{2}+y^{2}
\end{equation}
and that
\begin{equation}
x^{2}+y^{2}+z_{1}^{2}+z_{2}^{2}+r^{2}=1\,,\label{eq:Cons_}
\end{equation}
so one variable is effectively superfluous. The dark energy equation
of state and effective equation of state read respectively:
\begin{eqnarray}
w_{{\rm DE}} & = & \frac{x^{2}-y^{2}}{x^{2}+y^{2}}\,,\label{eq:EOS_DE}\\
w_{{\rm eff}} & = & x^{2}-y^{2}\,.\label{eq:EOS_eff}
\end{eqnarray}

We plot in Figs.~\ref{p_1mu_0} and \ref{p_1mu_08} the behavior of
the model for some selected values of the parameters. The top panels
display plots of $\Omega_{{\rm DE}}$, $\Omega_{-}$ and $\Omega_{+}$
for $\alpha=0.5,\,0.8,\,1.3,\,1.6$ and $\beta=1$. These values are
chosen to lie progressively farther from the SN best fit. Hence, as
expected, we see that the plots deviate more and more from $\Lambda$CDM.
The middle and bottom panels display the evolution of the dark energy
equation of state (EOS) $w_{{\rm DE}}$ and effective EOS $w_{{\rm eff}}$,
respectively, for the same parameter values. For $w_{{\rm DE}}$ and
$w_{{\rm eff}}$, we also plot the theoretical predictions for the
asymptotic behavior (see Section \ref{critical_points}). Notice that
after an initial transient, the plots become insensitive to $\mu_{{\rm in}}$,
as will be confirmed quantitatively with the likelihood analysis of
Section \ref{results}.

\section{Critical points}

\label{critical_points}

\begin{table*}
\begin{tabular}{|c|c|c|c|c|c|c|c|c|}
\hline
No  & $x$  & $y$  & $z_{1}$  & $z_{2}$  & \,$\Omega_{{\rm DE}}$\,  & $w_{\phi}$  & $w_{eff}$  & $\mu$\tabularnewline
\hline
\hline
1  & $\pm1$  & 0  & 0  & 0  & 1  & 1  & 1  & 0\tabularnewline
\hline
2  & $\frac{\alpha}{3}$  & $\frac{1}{3}\sqrt{9-\alpha^{2}}$  & 0  & 0  & 1  & $-1+\frac{2\alpha^{2}}{9}$  & \,\,$-1+\frac{2\alpha^{2}}{9}$\,\,  & 0\tabularnewline
\hline
3  & 0  & 0  & $\frac{1}{\sqrt{2}}$  & $\frac{1}{\sqrt{2}}$  & 0  & 0  & 0  & $0$\tabularnewline
\hline
4  & $-\frac{2\beta}{3}$  & 0  & $\sqrt{1-\frac{4\beta^{2}}{9}}$  & 0  & $\frac{4\beta^{2}}{9}$  & 1  & $4\frac{\beta^{2}}{9}$  & 1\tabularnewline
\hline
5  & \,\, $\frac{3}{2(\alpha+\beta)}$ \,\,  & \,\, $\frac{\sqrt{9+4\alpha\beta+4\beta^{2}}}{2|\alpha+\beta|}$\,\,  & \,\,$\frac{\sqrt{-9+2\alpha\beta+2\alpha^{2}}}{\sqrt{2}|\alpha+\beta|}$
\,\,  & 0  &\,\, $\frac{9+2\alpha\beta+2\beta^{2}}{2(\alpha+\beta)^{2}}$ \,\, & \,\, $\frac{-2(\alpha+\beta)\beta}{9+2\alpha\beta+2\beta^{2}}$ \,\, & \,\,$\frac{-\beta}{(\alpha+\beta)}$\,\,  & $1$\tabularnewline
\hline
\end{tabular}\caption{\label{tab:Full-Table-of}Critical points. Only points 2 and 5 can
have accelerated expansion ($w_{{\rm eff}}<-1/3$). See Section \ref{critical_points}
for more details.}
\label{tbl_crt} %
\end{table*}

The goal of this paper is the solution of the system of field equations
(\ref{eq:Frid_EQ-H-prime})-(\ref{eq:Frid_EQ-5}) and the comparison
with SN Ia data. Expressing $z_{2}$ in terms of $x,y,z_{1}$
via Eq.~(\ref{eq:Cons_}) and employing the fact that we are studying
the background evolution in the matter era (and so $r=0$), we can
reduce the system of equations by two. An important step is the analysis
of the critical points $x'=y'=z_{1}'=0$, i.e.~the solution of the
following system of equations: \begin{align}
0 & =\frac{x}{2}(3-3y^{2}+3x^{2})-3x+\alpha y^{2}\label{eq:Cr_Frid_EQ-1}\\
 & \phantom{po}+\beta(1-x^{2}-y^{2}-2z_{1}^{2})\,,\nonumber \\
0 & =\frac{y}{2}(3-3y^{2}+3x^{2})-\alpha xy\,,\label{eq:Cr_Frid_EQ-2}\\
0 & =\frac{z_{1}}{2}(3-3y^{2}+3x^{2})-\frac{3}{2}z_{1}+\beta xz_{1}\,.\label{eq:Cr_Frid_EQ-3}\end{align}
 We need to select the critical points which are real and physical,
i.e.~with real positive energy density less than unit. This implies
$|x|\le1$ and $0\le y,z_{1},z_{2}\le1$. Employing the symmetries under
$\beta\to-\beta$, $z_{1}\to z_{2}$, and under $\alpha\to-\alpha$,
$x\to-x$ and $\beta\to-\beta$, we can discard a number of symmetrical
points and restrict our analysis to $\alpha\ge0$, $\beta\ge0$, $\mu\ge0$.
All the surviving points are listed in Table \ref{tab:Full-Table-of}.
Since we need to have a final state that is accelerated ($w_{{\rm eff}}<-1/3$),
only points 2 and 5 can be acceptable final states. The stable acceleration
regions of points 2 and 5 are then found by linearly perturbing the
system around these points.

\begin{figure}
\includegraphics[width=1\columnwidth]{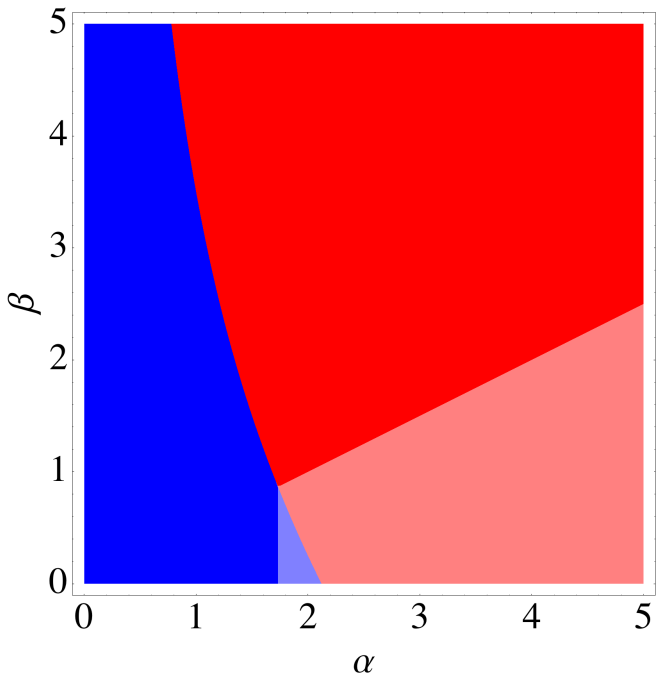} \caption{Regions of stability for point 2 (blue) and point 5 (red) of Table
\ref{tbl_crt}. The darker blue and red regions have a final state
that is accelerated ($w_{{\rm eff}}<-1/3$). In the following, we will
restrict to these regions of stable acceleration.}

\label{Cr_tot} %
\end{figure}

The eigenvalues of the linearization matrix around point 2 are: \begin{align}
\lambda_{1}^{p2} & =\frac{-9+\alpha^{2}}{3}\,,\nonumber \\
\lambda_{2}^{p2} & =\frac{-9+2\alpha^{2}-2\alpha\beta}{3}\,,\label{eq:Cond_crit-1}\\
\lambda_{3}^{p2} & =\frac{-9+2\alpha^{2}+2\alpha\beta}{6}\,,\nonumber \end{align}
 and its region of stability is defined by the condition Re$[\lambda_{1,2,3}]<0$.
The region of the parameter space that is stable and accelerated is
shown in dark blue in Fig.~\ref{Cr_tot}. We will restrict the analysis
of Sections \ref{supernovae} and \ref{results} to this region.

The eigenvalues for point 5 are: \begin{align}
\lambda_{1}^{p5} & =-\frac{6\beta}{\alpha+\beta}\,,\nonumber \\
\lambda_{2}^{p5} & =-\frac{3\alpha+6\beta+\Delta(\beta)}{4(\alpha+\beta)}\,,\label{eq:Cond_crit-1-1-1}\\
\lambda_{3}^{p5} & =\frac{-3\alpha-6\beta+\Delta(\beta)}{4(\alpha+\beta)}\,,\nonumber \end{align}
 where $\Delta(\beta)\equiv[9(-36+7\alpha^{2})+4\alpha(-27+8\alpha^{2})\beta+4(-45+16\alpha^{2})\beta^{2}+32\alpha\beta^{3}]^{1/2}$.
The region of stable acceleration for point 5 is shown in dark red
in Fig.~\ref{Cr_tot}.

\begin{figure}
\includegraphics[width=0.95\columnwidth]{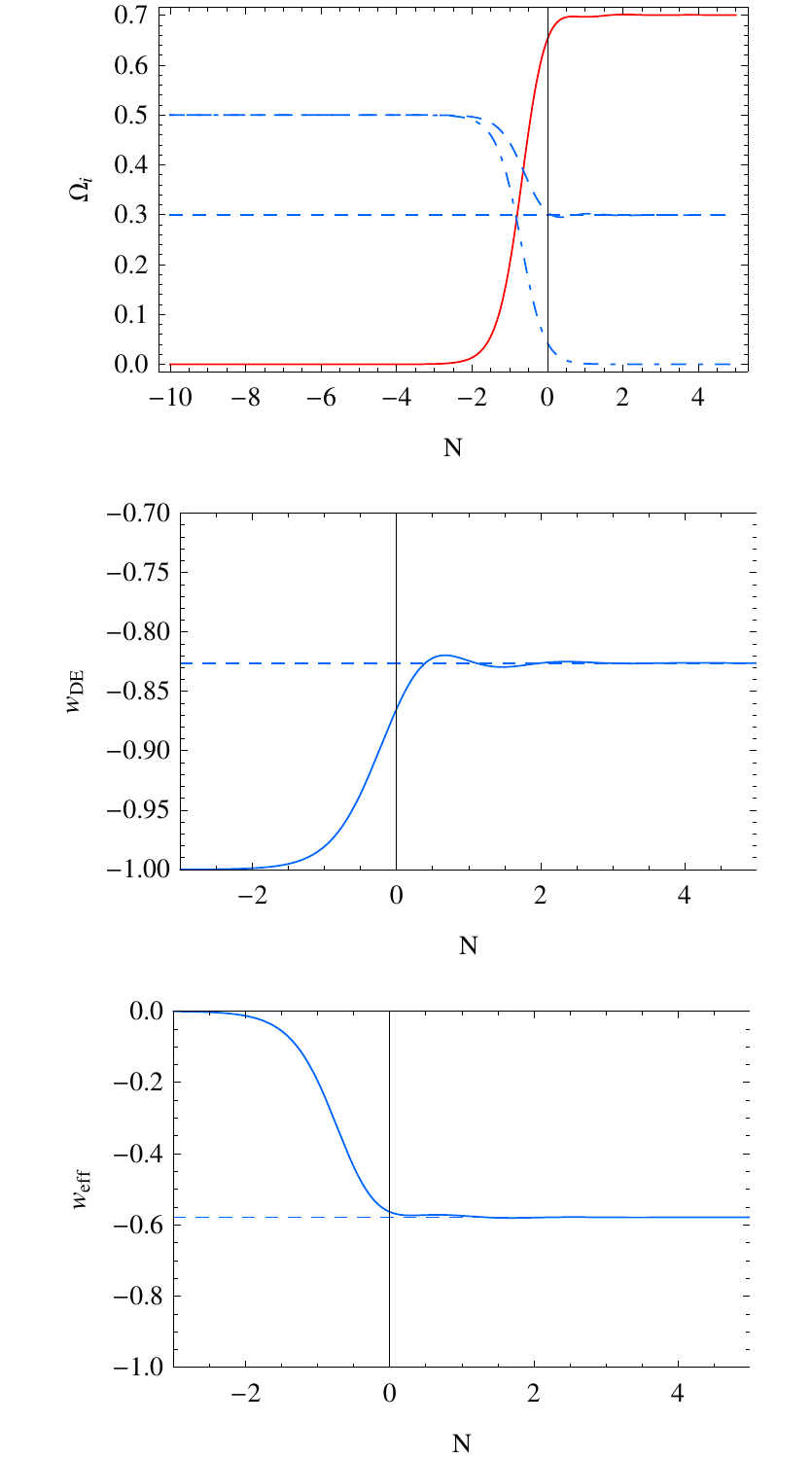}
\caption{In the top panel, $\Omega_{{\rm DE}}$ (solid, red), $\Omega_{-}$ (dot-dashed) and $\Omega_{+}$
(dashed) for $\beta=3.52$, $\alpha$= 2.56 and $\mu_{{\rm in}}$=
0. For these values, point 5 is stable and accelerated. The second
and the third panels show $w_{{\rm DE}}$ and $w_{{\rm eff}}$, respectively.
The long-dashed straight lines mark the asymptotic behaviors of the corresponding
quantities. Plots are with respect to the e-folding time variable
$N$.}

\label{p_87mu_0a1_p6} %
\end{figure}

Point 5 allows for a non-vanishing asymptotic value of $\Omega_{\pm}$,
while still producing acceleration. This is illustrated in Fig.~\ref{p_87mu_0a1_p6}.
Point 5 is interesting since it automatically selects one of the two
forms of matter to dominate asymptotically along with dark energy,
while the other one vanishes. This type of solutions, dubbed scaling
or stationary solutions (see e.g. \cite{2006PhRvD..74b3525A}), have
been found in many other systems of interacting models.

\section{SN Ia data and method}

\label{supernovae}

In order to constrain the parameters of the model discussed so far,
we will use the Union2.1 Compilation~\cite{Suzuki_etal_2012} of
580 Type Ia SN in the redshift range $z=0.015-1.414$.\footnote{ More precisely, we use the magnitude vs.~redshift table publicly available at the Supernova Cosmology Project \href{http://supernova.lbl.gov/Union/}{webpage}. This table does not include systematic errors.}
The predicted magnitudes are related to the luminosity distance $d_{L}$ by: \begin{equation}
m(z)=5\log_{10}\frac{d_{L}(z)}{10\,\textrm{pc}}\,,\end{equation}
 which is computed assuming spatial flatness: \begin{equation}
d_{L}(z)=(1+z)\int_{0}^{z}\frac{\rd\bar{z}}{H(\bar{z})}\,.\end{equation}
 The likelihood analysis is based on the $\chi^{2}$ function: \begin{equation}
\chi_{SNIa}^{'2}=\sum_{i}\frac{[m_{i}-m(z_{i})+\xi]^{2}}{\sigma_{i}^{2}}\,,\end{equation}
 where the index $i$ labels the elements of the Union2.1 dataset.
The parameter $\xi$ is an unknown offset sum of the supernova absolute
magnitudes, of $k$-corrections and other possible systematics. As
usual, we marginalize the likelihood $L'_{SNIa}=\exp(-\chi_{SNIa}^{'2}/2)$
over $\xi$, $L_{SNIa}=\int\rd\xi\, L'_{SNIa}$, leading to a new marginalized
$\chi^{2}$ function: \begin{equation}
\chi_{SNIa}^{2}=S_{2}-\frac{S_{1}^{2}}{S_{0}}\,,\end{equation}
 where we neglected a cosmology-independent normalizing constant,
and the auxiliary quantities $S_{n}$ are defined as: \begin{equation}
S_{n}\equiv\sum_{i}\frac{\left[m_{i}-m(z_{i})\right]^{n}}{\sigma_{i}^{2}}\,.\end{equation}
 As $\xi$ is degenerate with $\log_{10}H_{0}$, we are effectively
marginalizing also over the Hubble constant.

The luminosity distance $d_{L}(z)$ is obtained by integrating numerically
Eqs~(\ref{eq:Frid_EQ-H-prime})-(\ref{eq:Frid_EQ-5}). For every
value of the parameters $\alpha,\beta,\mu_{{\rm in}}$ and every possible
value of $\Omega_{{\rm DE},0}$ we begin with a trial initial condition
at a very large initial $z$ that is given by $\Lambda$CDM with that
particular $\Omega_{{\rm DE},0}$ and with $\Omega_{{\rm m},0}=\Omega_{+0}+\Omega_{-0}=1-\Omega_{{\rm DE},0}$.
At $z=0$ this solution will produce a $\Omega_{{\rm DE},0}$ different
from the $\Lambda$CDM value. We then perturb by trial and error the
initial values of $x,y,z_{1},z_{2}$ until we find the sought-for
$\Omega_{{\rm DE},0}$. We always impose as initial value $x=0$.

\section{Results}

\label{results}

We will now show how the Union2.1 SN Compilation constrains the McDE
scenario discussed in this paper. As discussed in Section \ref{critical_points},
we will consider only positive values of $\{\alpha,\beta,\mu_{{\rm in}}\}$
because of symmetry. Furthermore, we will analyze separately the regions
of stable acceleration of points 2 and 5 (dark blue and red in Fig.~\ref{Cr_tot},
respectively). The reason is that the attractor of point 2 depends
only on $\alpha$, while the attractor of point 5 depends on both
$\alpha$ and $\beta$, as shown in Table \ref{tbl_crt}. Because
of this degeneracy between $\alpha$ and $\beta$, it is more instructive
to keep the analysis of point 5 separate from the one of point 2.

\subsection{Constraints for critical point 2}

\begin{table}[!ht]
 \begin{ruledtabular} \begin{tabular}{ccccc}
Parameter  & Best fit  & 95\% c.l.  &  & \tabularnewline
\hline
$\Omega_{{\rm DE},0}$  & 0.719  & $[0.680,0.765]$  &  & \tabularnewline
$\alpha$  & 0.62  & $[0,1.01]$  &  & \tabularnewline
$\beta$  & 6.4  & $[0,83]$  &  & \tabularnewline
$\mu_{{\rm in}}$  & unconstrained  & unconstrained  &  & \tabularnewline
\hline
\end{tabular}\end{ruledtabular} \caption{Best-fit values with 95\% confidence intervals for the parameters
of the model discussed in this paper. We employed a flat prior on
the parameters, and in particular $0\le\beta\le100$ and we explore
here only the point 2 region of stability and acceleration. See Fig.~\ref{1dpost}
for a plot of the marginalized posterior distributions.}

\label{tab:1D} %
\end{table}

In order to constrain point 2 with the Union2.1 SN Compilation we
compute the likelihood $L_{{\rm SNIa}}$, discussed in the previous Section,
on a grid in the 4-dimensional parameter space $\{\Omega_{{\rm DE},0},\alpha,\beta,\mu_{{\rm in}}\}$
and we marginalize analytically over the absolute magnitude of the
supernovae.

\begin{figure*}
\includegraphics[width=0.9\columnwidth]{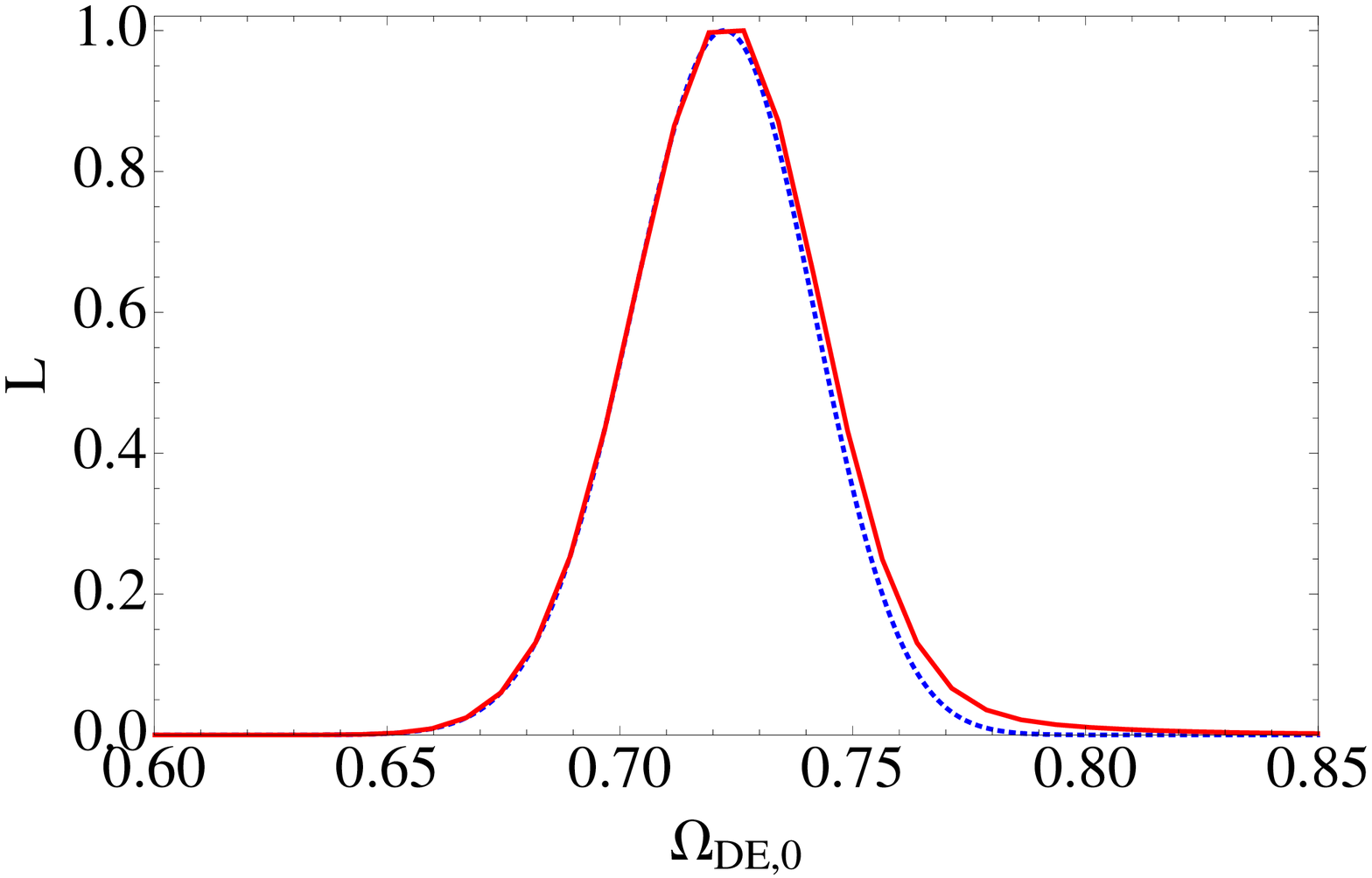} \qquad{}\includegraphics[width=0.9\columnwidth]{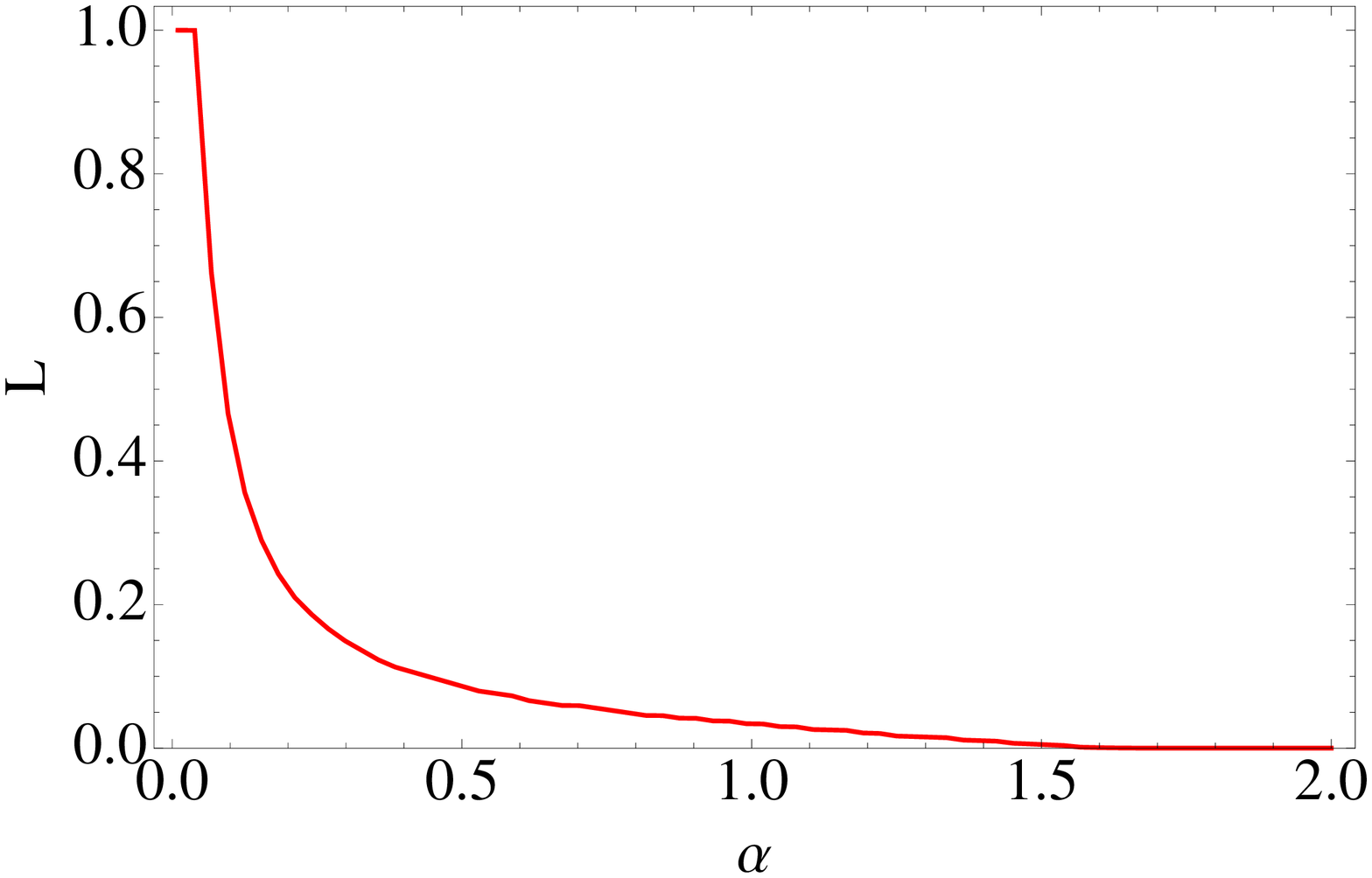}
\hspace{.9\columnwidth} \phantom{ciao} \\
 \includegraphics[width=0.9\columnwidth]{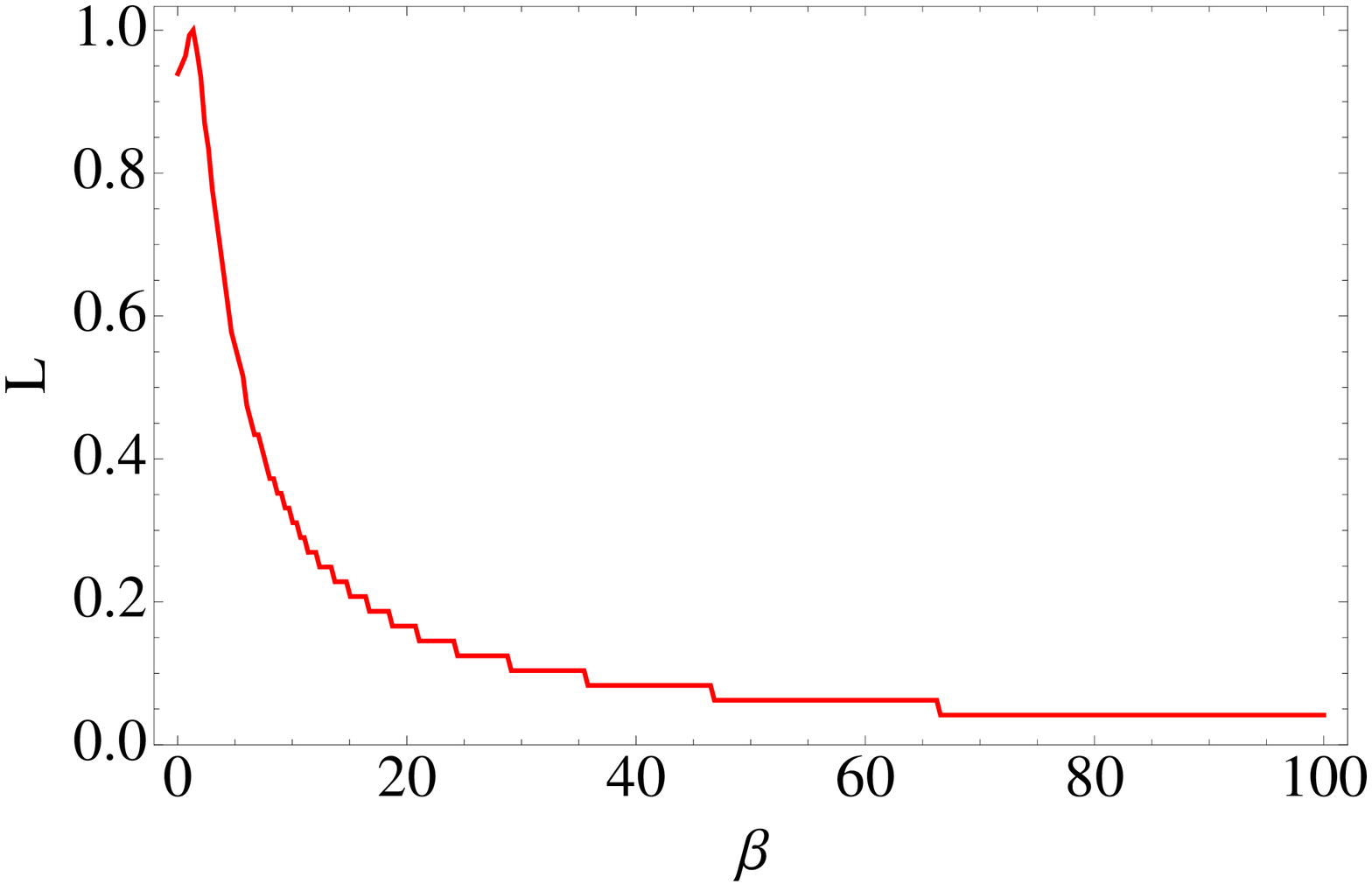} \qquad{}\includegraphics[width=0.9\columnwidth]{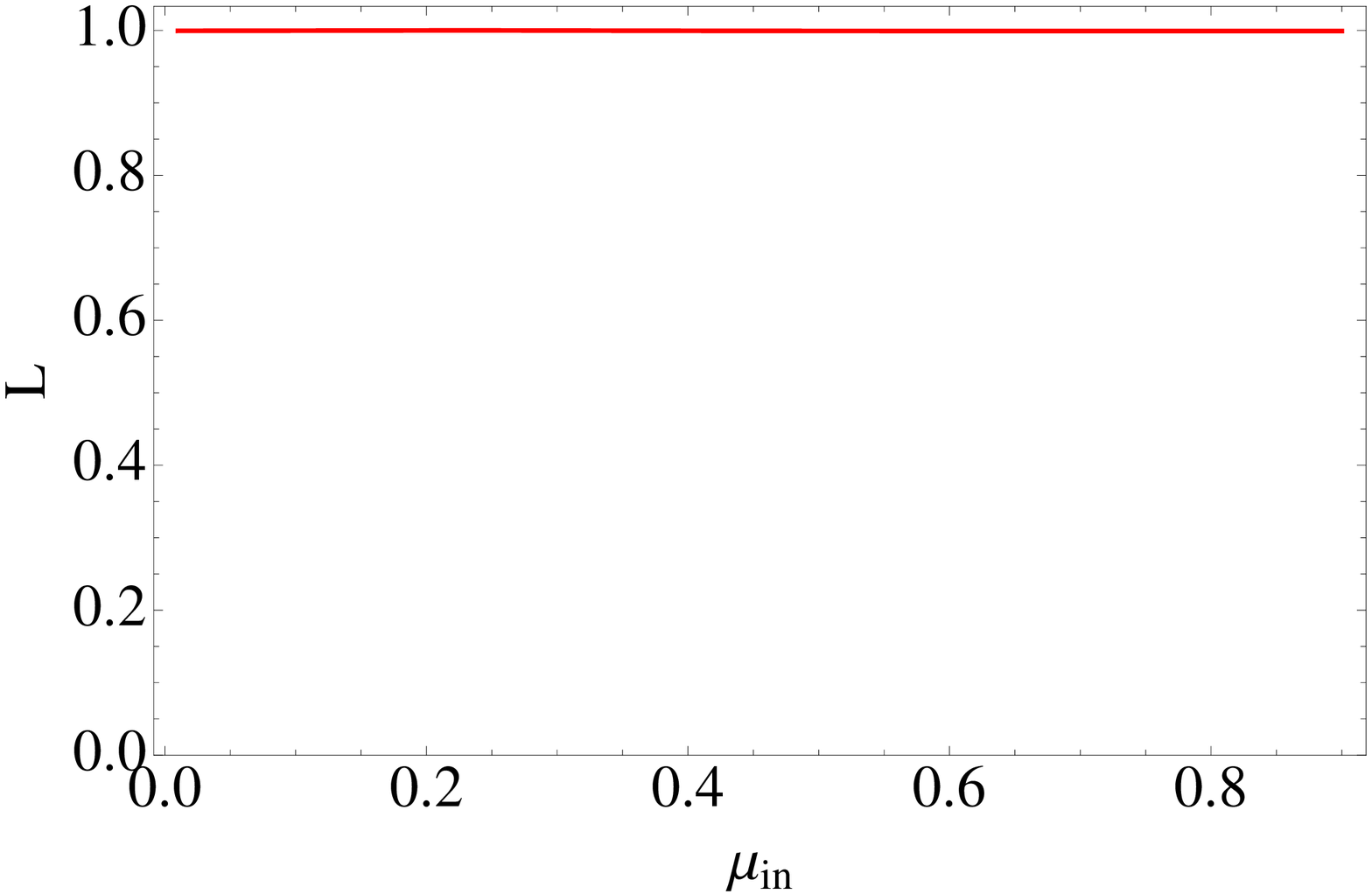}
\caption{Marginalized posterior distributions for point 2 on the parameters
$\{\Omega_{{\rm DE},0},\alpha,\beta,\mu_{{\rm in}}\}$ when fitting the
model of this paper to the Union2.1 SN Compilation~\cite{Suzuki_etal_2012}.
In the top-left panel the dotted curve is the posterior of the flat
$\Lambda$CDM model. While SNIa observations constrain to a finite range
of values both the present-day amount of dark energy $\Omega_{{\rm DE},0}$
and potential slope $\alpha $, they can only marginally constrain
the coupling $\beta$ and not at all the value of the asymmetry parameter
$\mu_{{\rm in}}$. See Table~\ref{tab:1D} for best-fit values with
95\% confidence intervals. }

\label{1dpost} %
\end{figure*}

Fig.~\ref{1dpost} shows the marginalized posterior distributions
on the single parameters. The best-fit values with 95\% confidence
levels are shown in Table~\ref{tab:1D}. From these results, it is
clear that  SNIa can constrain to a finite range both the present-day
amount of dark energy $\Omega_{{\rm DE},0}$ and the exponent $\alpha$
of the potential. However, background observables such as SNIa can
only marginally constrain the coupling $\beta$ (the likelihood stays
constant for values $\beta>100$) and cannot constrain at all the
value of the asymmetry parameter $\mu_{{\rm in}}$. In the top-left
panel of Fig.~\ref{1dpost} we also show the posterior of $\Omega_{{\rm DE},0}$
for the case of the flat $\Lambda$CDM model (dotted curve). As one
can see, the constraints are only slightly weaker for larger values
of the dark energy density. Furthermore, the minimum $\chi^{2}$ for
the two models -- McDE and $\Lambda$CDM -- is approximately the same,
$\chi_{{\rm min}}^{2}=562.2$.

\begin{figure}
\includegraphics[width=1\columnwidth]{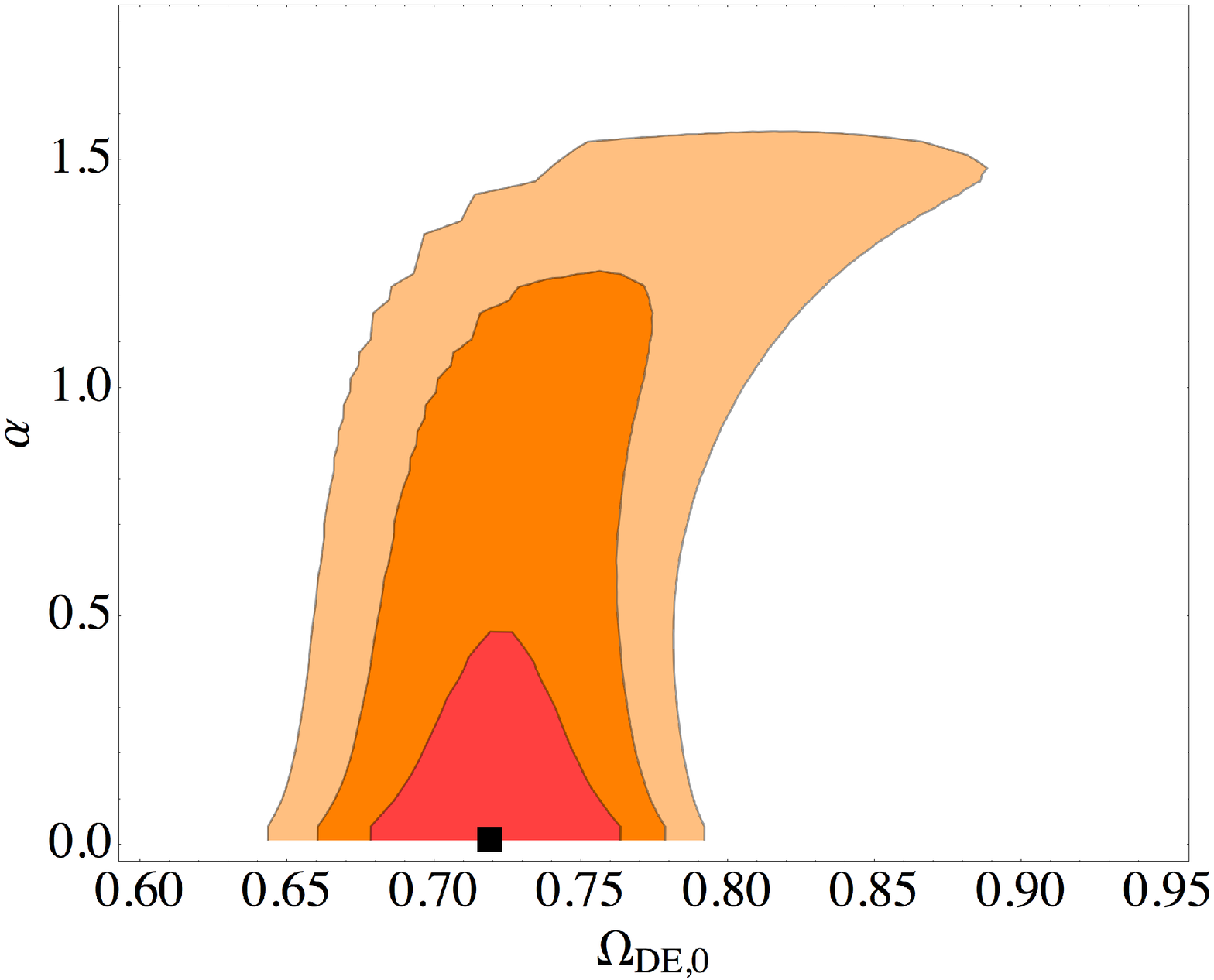}\\
 \ \\
 \includegraphics[width=1\columnwidth]{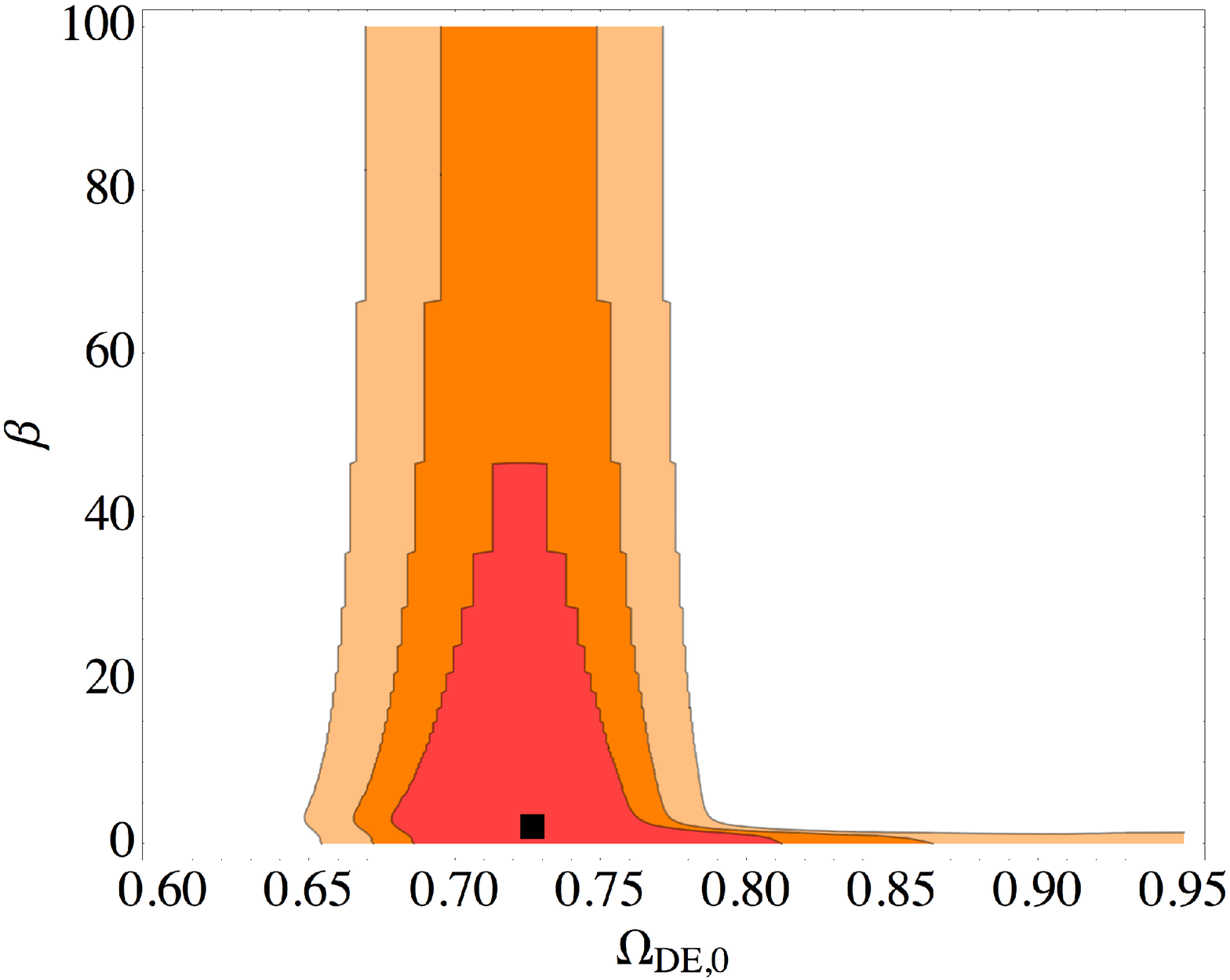}\\
 \ \\
 \includegraphics[width=1\columnwidth]{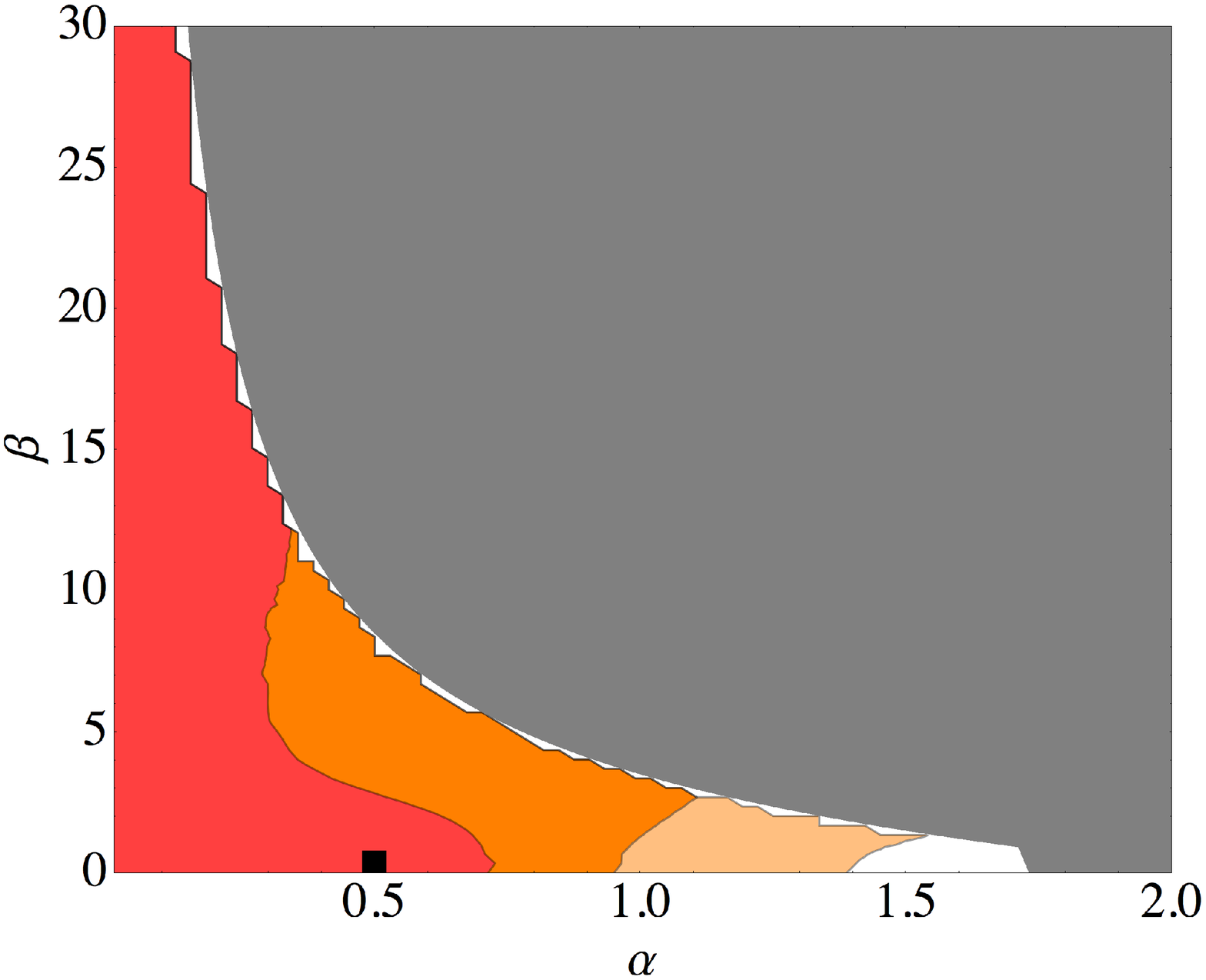} \caption{ $1\sigma$, $2\sigma$ and $3\sigma$ confidence-level contours for
the relevant 2-dimensional marginalized posterior distributions for
point 2. The black squares mark the best-fit values. The degeneracy
between $\Omega_{{\rm DE},0}$ and the parameters $\alpha,\beta$ makes values
of $\Omega_{{\rm DE},0}\simeq0.90$ possible at the $3\sigma$ level. The
bottom panel explains why the best fit has $\alpha=0.62$, while the
posterior on $\alpha$ is peaked at $\alpha=0$. The gray region is
excluded, see Fig.~\ref{Cr_tot}.}

\label{2dpostP2} %
\end{figure}

By comparing the best-fit values reported in Table~\ref{tab:1D}
with the maximum of the posteriors shown in Fig.~\ref{1dpost}, one
can notice a mismatch. To investigate this issue -- and also the level
of degeneracy between the parameters -- we show in Fig.~\ref{2dpostP2}
the relevant 2-dimensional marginalized posterior distributions. One
can see, perhaps a little surprisingly, that values of $\Omega_{{\rm DE},0}\simeq0.90$
are allowed at the $3\sigma$ level; this is clearly due to the degeneracy
between $\Omega_{{\rm DE},0}$ and the parameter $\alpha$ when $\beta$
is negligible and it is similar to the well-known degeneracy between
$\Omega_{{\rm DE},0}$ and $w$ for a dark energy model with constant EOS. Also, the posterior on $\alpha$ and $\beta$ shows a nontrivial
likelihood surface. This explains the mismatch between the best-fit
values of Table~\ref{tab:1D} and the maximum of the posterior on
$\alpha$.

\subsection{Constraints for critical point 5}

\begin{figure}
\includegraphics[width=1\columnwidth]{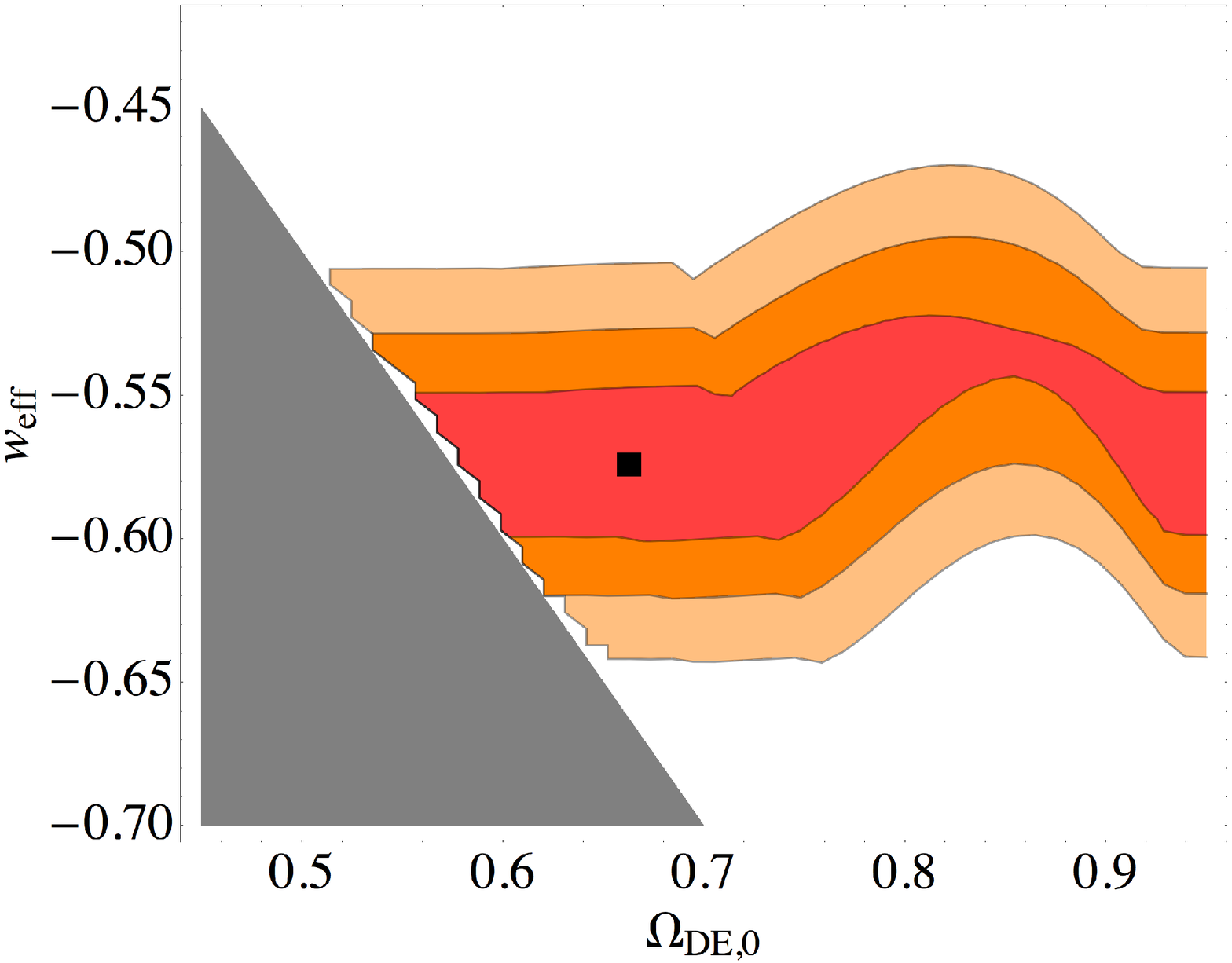}\\
 \ \\
 \includegraphics[width=1\columnwidth]{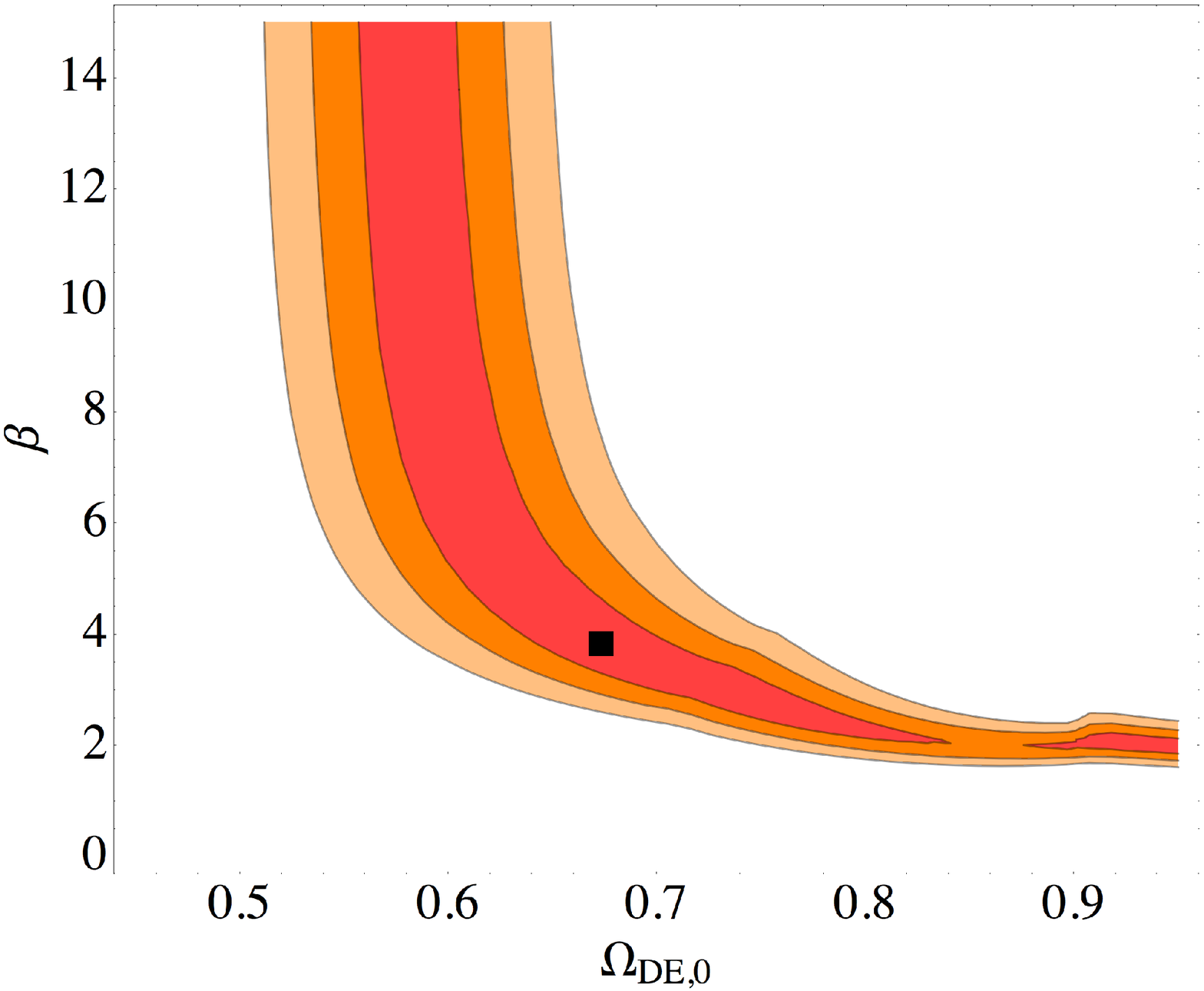}\\
 \ \\
 \includegraphics[width=1\columnwidth]{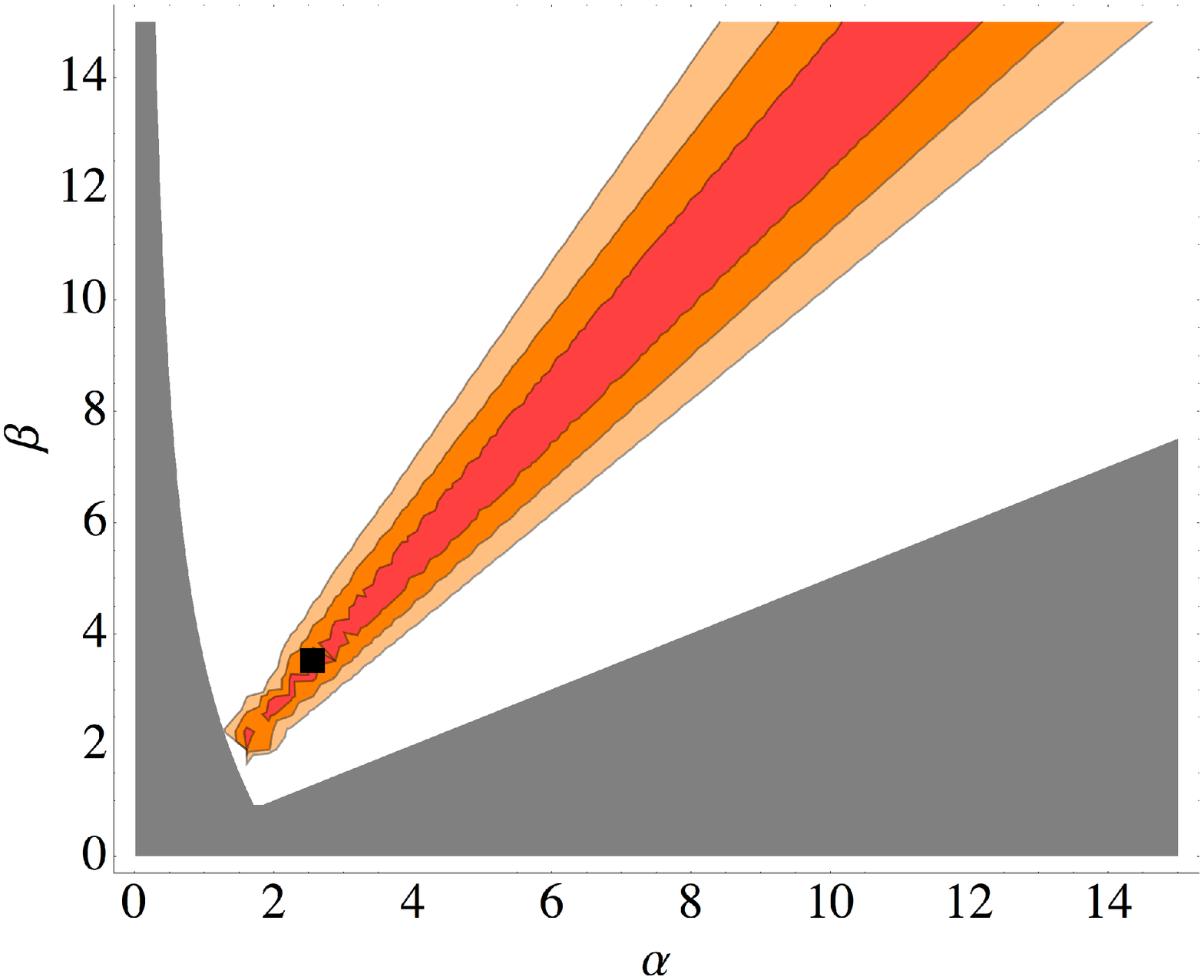} \caption{ $1\sigma$, $2\sigma$ and $3\sigma$ confidence-level contours for
the 2-dimensional marginalized posterior distributions relative to
asymptotic solutions of point 5. The black squares mark the best-fit
values. The gray regions are excluded, see Fig.~\ref{Cr_tot} and
Table~\ref{tab:Full-Table-of}. For each posterior, a flat prior on the parameters
displayed in the plot has been adopted. Particularly evident is the
degeneracy between $\alpha$ and $\beta$.}

\label{2dpostP5} %
\end{figure}

 As anticipated and shown in Table~\ref{tab:Full-Table-of}, the attractor of
critical point 5 depends on both the potential slope $\alpha$
and the coupling $\beta$. Therefore, a strong degeneracy between
$\alpha,\beta$ is to be expected. So as to better understand the
behaviors of the solutions relative to point 5, we will then restrict
to trajectories for which the asymptotic attractor has already been
reached at the present time. If on one hand this simplifies and clarifies
the analysis, on the other hand this choice worsens the overall fit
($\chi_{{\rm min}}^{2}=568.2$). This is to be expected as the effective
EOS $w_{{\rm eff}}$ is now constant, and so clearly
in tension with SN data. We checked that non-asymptotic solutions
relative to point 5 can give as good a fit ($\chi_{{\rm min}}^{2}=562.4$)
as point 2 and the flat $\Lambda$CDM model.

Fig.~\ref{2dpostP5} shows the constraints from SNIa observations for
the case of asymptotic solutions of point 5 for three pairs of parameters.
The top panel shows that values of the (constant) effective EOS $w_{{\rm eff}}$ quite larger than $-1$ are preferred. This
is to be expected as SNIa prefer models with $w_{{\rm eff}}$ evolving
from 0 to $\sim-1$ in the future, with a present value around $w_{{\rm eff}}\approx w\Omega_{{\rm DE},0}\approx-\Omega_{{\rm DE},0}$.
The second plot shows a nontrivial degeneracy between $\Omega_{{\rm DE},0}$
and $\beta$, while the bottom plot shows the expected very strong
degeneracy between $\alpha$ and $\beta$. From Table \ref{tbl_crt}
one can in fact see that for large $\alpha,\beta$ point 5 depends
only on $\beta/\alpha$. The best-fit values for the three analyses are $(\Omega_{DE},w_{eff})=(0.66,-0.57)$, $(\Omega_{DE}, \beta)=(0.67,3.82)$, $(\alpha,\beta)=(2.56,3.52)$, respectively.
\ \\

Summarizing, we found that the proposed model is compatible with SNIa
observations in a large range of parameters. In particular, we find
that the cosmic expansion is very weakly dependent on the dimensionless
coupling constant $\beta$ and even values of $\sim 100$ (i.e. three orders of magnitude larger than present bounds on the coupling $\beta \lesssim 0.1$ for standard cDE models)
are allowed. Moreover, we found that the luminosity distance probed
by SNIa data is insensitive to $\mu_{{\rm in}}$. We have repeated the
analysis of this Section using the forecasted Dark Energy Survey (DES)
sample of 3000 SNe~\cite{Bernstein:2011zf} and found that while the constraints on $\Omega_{DE}$ become tighter, the constraints on the the McDE parameters $\alpha, \beta$ are only marginally improved.
This further demonstrates the ability of McDE to satisfy background observations.

We expect, however, that the value of $\mu_{{\rm in}}$ will be very
important as far as structure formation is concerned and that the
analysis of perturbations will bring improved constraints. In fact,
as shown in \citep{Baldi_2012a}, the perturbation equations of McDE
cosmologies are unstable with respect to the asymmetry between the
two different CDM species, due to the opposite signs of the extra-friction
terms \citep[see][for a more detailed discussion on the instability of adiabatic perturbations in McDE models]{Baldi_2012a}.
As a consequence, the growing asymmetry at low redshifts related to
the dynamical evolution of the background is expected to amplify the
effects of the attractive and repulsive fifth-forces at the level
of linear and nonlinear density perturbations \citep[see again][]{Baldi_2012c}.

\section{Conclusions}

\label{conclusions}

In the present paper, we have compared for the first time the Multi-coupled
Dark Energy scenario proposed by \cite{Baldi_2012a} with observational
data on the cosmic expansion history, consisting of the 580 Type Ia
supernovae of the Union2.1 sample~\cite{Suzuki_etal_2012}.\\

As a first step, we have performed a full analytic study of the dynamical
system associated to the background equations of Multi-coupled Dark
Energy, identifying the critical points of the system and investigating
in full detail their existence and stability conditions. Among the
physically consistent points, we have selected those giving cosmologically
viable solutions. Such analysis has shown that all critical points
are characterized by only three possible asymptotic values of the
asymmetry parameter $\mu$ that quantifies the relative abundance
of the two CDM particle species. Such values are $0$ and $\pm1$
(due to symmetry between $\mu\to-\mu$), corresponding to a perfect
balance between the two species and to the full domination of one
of the two species over the other, respectively. This means that,
independently of the initial value of the asymmetry parameter $\mu$,
the asymptotic fate of the universe in McDE cosmologies will be characterized
by either perfect symmetry between the two CDM species, or by the
disappearance of one of the two. Additionally, our results identified
a nontrivial solution for the future evolution of the universe characterized
by a scaling between the Dark Energy and CDM relative densities that
keep a constant ratio in the asymptotic future.

As a second step, we have performed a full likelihood analysis in
the model's parameter space, restricting the range of parameters based
on the stability conditions previously obtained. As expected, we find
that supernova data can constrain the slope of the self-interaction
potential $\alpha$, which is found to be bound to values $\lesssim1.5$
at the $3\sigma$ confidence level. On the other hand, we find a flat
posterior likelihood for the initial asymmetry parameter, $\mu_{{\rm in}}$,
which is therefore completely unconstrained by the data, and more
interestingly we derive the $1$, $2$, and $3\sigma$ confidence
regions in the parameter plane $\beta-\Omega_{{\rm DE}\,,0}$, showing very
weak constraints on the coupling parameter $\beta$. In fact, couplings
of the order of $\beta\sim10^{2}$ appear to be consistent at the
$2\sigma$ confidence level with the fiducial value of the Dark Energy
fractional density. The correction to the gravitational force is proportional
to $\beta^{2}$; this implies that an extra force, four orders of magnitude
stronger than gravity, would still go unnoticed in a SN Hubble diagram.

The results of our direct comparison show that the Multi-coupled Dark
Energy scenario is practically indistinguishable from a standard $\Lambda$CDM
cosmology at the background level, thereby providing a direct example
of how the background expansion history might be almost insensitive
to the internal complexity of the dark sector. These conclusions still
hold if one uses the forecasted DES sample of 3000 SNIa instead of
the Union2.1 SN Compilation.

Observational constraints on the linear growth of density perturbations,
and ultimately the detailed investigation of the impact of Multi-coupled
Dark Energy models on the dynamical evolution of nonlinear structures
at small scales, promise to place much tighter bounds on the model's
parameter space, and will be investigated in future works. For now,
our investigation demonstrates the full viability of Multi-coupled
Dark Energy as an alternative to the $\Lambda$CDM cosmology for what
concerns the cosmic expansion history.
\begin{acknowledgments}
A.P.~is grateful for the hospitality of the ITP, University of Heidelberg,
where most of this work was prepared. A.P.~acknowledges funding by
DAAD for financial support. M.B.~is supported by the Marie Curie
Intra European Fellowship {}``SIDUN'' within the 7th Framework Programme
of the European Community and also acknowledges partial support by
the DFG Cluster of Excellence {}``Origin and Structure of the Universe''. L.A., M.B.~and V.M.~acknowledge partial support from the TRR33
Transregio Collaborative Research Network {}``The Dark Universe''.
\end{acknowledgments}
\bibliographystyle{mnras}
\bibliography{baldi_bibliography,bibliography_2,amendola,bibliography}

\end{document}